\documentclass{article}

\usepackage{spconf,amsmath,amsfonts,graphicx,hyperref,mathtools}
\usepackage{tikz}
\usepackage{algpseudocode, algorithmicx}
\newcounter{lemma}
\newenvironment{lemma}{\refstepcounter{lemma}
\noindent 
\textbf{Lemma \thelemma.} \em \rmfamily}{}

\newenvironment{lemma*}[1]{\refstepcounter{lemma}
\noindent 
\textbf{Lemma \thelemma~(#1)} \em \rmfamily}{}

\newcounter{theorem}
\newenvironment{theorem}[1]{\refstepcounter{theorem}
\noindent 
\textbf{Theorem \thetheorem~(#1)} \em \rmfamily}{}

\newcounter{assumption}
\newenvironment{assumption}{\refstepcounter{assumption}
\noindent 
\textbf{Assumption \theassumption.} \em \rmfamily}{}

\newcounter{proposition}
\newenvironment{proposition}{\refstepcounter{proposition}
\noindent 
\textbf{Proposition \theproposition.} \em \rmfamily}{}

\newcounter{problem}

\newenvironment{problem*}[1]{\refstepcounter{problem}
\noindent 
\textbf{Problem \theproblem~(#1)} \em \rmfamily}{}

\newcounter{remark}
\newenvironment{remark}{\refstepcounter{remark}
\noindent 
\textit{Remark \theremark.} \em \rmfamily}{}

\newcommand{\ind}[1]{\mathbb{I}_{\{#1\}}}
\usepackage{algorithm}
\usepackage{enumitem}

\DeclareMathOperator*{\argmax}{arg\,max}

\title{Achieving Pareto Optimality in Games via Single-bit Feedback}
\name{Seref Taha Kiremitci, Ahmed Said Donmez, Muhammed O. Sayin }
\address{Department of Electrical and Electronics Engineering, Bilkent University, Ankara, T\"{u}rkiye}
\begin{document}
%
\maketitle
\begin{abstract}
Efficient coordination in multi-agent systems often incurs high communication overhead or slow convergence rates, making scalable welfare optimization difficult. We propose \emph{Single-Bit Coordination Dynamics for Pareto-Efficient Outcomes (SBC-PE)}, a decentralized learning algorithm requiring only a single-bit satisfaction signal per agent each round. Despite this extreme efficiency, SBC-PE guarantees convergence to the \emph{exact} optimal solution in arbitrary finite games. We establish explicit regret bounds, showing expected regret grows only logarithmically with the horizon, i.e., $O(\log T)$. SBC-PE combines minimal signaling, general applicability, and finite-time guarantees. These results show scalable welfare optimization is achievable under minimal communication constraints.
\end{abstract}
\begin{keywords}
Pareto efficiency, single-bit communication, regret analysis, multi-agent system
\end{keywords}

\section{Introduction}
Achieving socially optimal coordination in multi-agent systems is challenging under communication constraints. In applications such as wireless sensor networks or dynamic spectrum access, agents seek to optimize a global objective (social welfare) using only local observations and minimal feedback. Prior work has shown that even payoff-based learning rules can steer agents toward efficient outcomes in generic $n$-player games without inter-agent communication~\cite{marden6426834,doi:10.1137/110850694,PRADELSKI2012882}. However, these approaches typically lack effective finite-time performance guarantees. This limitation motivates \emph{communication-efficient} solutions with sharp convergence guarantees.


Related work examines welfare maximization in a wind farm setting, where agents seek to maximize their combined power output despite unknown interactions and have access to total farm power through communication \cite{marden2013wind}. The authors use a model-free randomized experimentation scheme with this aggregate feedback and establish asymptotic optimality guarantees for discretized controls under fixed wind conditions. Menon and Baras~\cite{menon2013distributed} address welfare optimization using local payoff observations and binary messages, without communicating numerical payoff values or total welfare. Their algorithm adds communication of agents' internal moods, which influence subsequent action choices, to earlier payoff-based learning dynamics \cite{marden6426834}, replacing the interdependence restriction with a connectivity condition on the combined interaction and communication graphs. They provide an asymptotic convergence guarantee under this condition and a suitable decreasing exploration schedule.

We propose \emph{Single-Bit Coordination Dynamics for Pareto-Efficient Outcomes (SBC-PE)}, a multi-agent learning algorithm that requires only one-bit broadcasting per agent per round. The method applies to arbitrary normal-form games (or distributed optimization problems) and aims to maximize a weighted sum of local payoffs. SBC-PE follows an explore-then-commit paradigm: agents initially explore by randomly sampling actions and subsequently commit to those local actions most frequently aligned with collective satisfaction. At each iteration, agents broadcast a single-bit signal generated probabilistically from their realized utilities and algorithm parameters. This minimal feedback constitutes the only exchanged information. With one-bit feedback per agent, SBC-PE provides an explicit finite-time guarantee for identifying the exact socially optimal outcome under the stated assumptions. This low communication burden makes SBC-PE particularly well suited for environments with bandwidth limitations or intermittent connectivity.


Communication efficiency is a central challenge in distributed learning, and strategies such as event-triggering, sparsification, and quantization have been proposed to reduce bandwidth and energy costs~\cite{com-eff-dsit-learn10038471}. Among these, quantization-based methods are especially appealing, with single-bit protocols representing the most communication-efficient extreme. Prior studies show that compressive and reduced-dimension diffusion strategies can achieve near full-information performance while significantly reducing communication overhead~\cite{reducedcomm6878433,singlebit6558808}. Similarly, consensus and optimization algorithms based solely on the \emph{sign} of relative state differences demonstrate that one-bit messages can suffice~\cite{discrete-sign8558107,CHEN20111962}, although such methods may require neighbor state estimates to determine the sign—an issue partially alleviated by compressed communication schemes~\cite{CAO2023111186}. Different from them, our contribution establishes that in arbitrary finite games, a single satisfaction bit per agent is sufficient to maintain communication efficiency while guaranteeing exact convergence to the socially optimal outcome with logarithmic regret bounds.


SBC-PE converges to the exact optimum in a fully decentralized manner under mild assumptions. For sufficiently large horizon $T$, the expected cumulative regret scales at most logarithmically as $O(\log T)$, with a precise characterization given in Theorem \ref{thm:main}. Both this theoretical result and our numerical experiments demonstrate that SBC-PE efficiently identifies the welfare-maximizing joint action in finite time. In contrast, earlier distributed learning methods emphasize coordination efficiency but often lack explicit performance quantification: for instance,~\cite{got10.1287/moor.2020.1051} achieves coordination without communication but is analyzed only asymptotically and is restricted to collusion-based allocation. While payoff-based learning rules can be stochastically stable at Pareto-efficient~\cite{marden6426834} or optimal equilibria~\cite{PRADELSKI2012882}, they typically rely on stronger structural assumptions or do not provide finite-time optimality guarantees. SBC-PE instead delivers explicit finite-time guarantees with minimal signaling.


Another notable advantage of SBC-PE is its \emph{generality}. The algorithm is agnostic to game structure and does not require assumptions such as potential games or specific reward distributions. This universality makes it applicable to a wide range of domains, including cooperative sensor coverage, multi-robot coordination, and interference management in dynamic spectrum access. In all such settings, agents autonomously converge to socially optimal configurations with negligible communication overhead.

The paper is organized as follows. In Sections \ref{sec:problem} and \ref{sec:algorithm}, we describe the problem and the SBC-PE algorithm. We present the analytical and numerical results in Sections \ref{sec:results} and \ref{sec:simulation} while provide concluding remarks in Section \ref{sec:conclusion}.




\section{Problem Formulation}
\label{sec:problem}

We consider a multi-agent game with agents $N = \{1,\dots,n\}$. Each agent $i$ selects an action $a_i \in A_i$, where $A_i$ is its finite action set. The joint action is $a = (a_1,\dots,a_n) \in A = \prod_{i=1}^n A_i$. Each agent $i$ then receives a utility $u_i(a)$. Agents collectively aim to maximize the social welfare:
\begin{equation}\label{eq:socialwelfare}
    W(a) = \sum_{i=1}^n w_i \cdot u_i(a),
\end{equation}
where $w_i > 0$ is the weight of agent $i$. Each agent can have local constraints, e.g., tolerable utility threshold $\lambda_i$, requiring
\begin{equation}
\label{eq: constraint}
    u_i(a) > \lambda_i, \quad \forall i \in N.
\end{equation}

Across $T$ stages, agents update their strategies depending only on the local realized utility and one-bit feedback from others. This enforces a decentralized, payoff-based learning rule where agents cannot observe actions or utilities of others, yet achieve efficient outcomes with minimal communication. 

\section{Algorithm}
\label{sec:algorithm}
\begin{algorithm}[tb]
    \caption{Single-Bit Coordination Dynamics for Pareto-Efficient Outcomes (SBC-PE)}
    \label{alg:sbc-pe}
    \begin{algorithmic}
    \State \textbf{Require:} exploration length $K$, local weight $w_i>0$, local threshold $\lambda_i$, common parameter $\epsilon \in (0,1)$
    \State \textbf{Initialize:} local counter $c_i(a_i)=0 \ \forall a_i \in A_i$
    \vspace{-0.4cm}
    \Statex
    \tikz[remember picture,overlay] {
        \node[rotate=90,anchor=south,yshift=5pt] at (0.2,-3) {\scriptsize \textbf{Exploration}};
        \draw[thick] (0,-5.5) -- (0,-0.3); 
    }
    \State \hspace{0.5em}\textbf{for} each time step $t=1,2,\ldots,K$ \textbf{do}
        \State \hspace{2.5em}choose action $a_i^t\sim\textrm{Uniform}(A_i)$
        \State \hspace{2.5em}observe local utility $u_i^t = u_i(a^t)$
        \State \hspace{2.5em}construct message
        \begin{equation}
            \label{eq:message}
            m_i^t =
            \begin{cases}
                \mathbb{I}_{\{u_i^t > \lambda_i\}} & \text{with probability } \epsilon^{1 - w_i \cdot u_i^t}, \\
                0 & \text{otherwise}.
            \end{cases}
         \end{equation}
        \State \hspace{2.5em}broadcast $m_i^t\in \{0,1\}$ to all other agents
        \State\hspace{2.5em}\textbf{if} $m_j^t = 1 \ \forall j \in N$ \textbf{then}
            \State \hspace{4.5em}increment $c_i(a_i^t) = c_i(a_i^t) + 1$
        \State \hspace{2.5em}\textbf{end if}
    \State \hspace{0.5em}\textbf{end for}
    \vspace{0.1cm}
    \State \hspace{0.5em}identify $\bar{a}_i \in \argmax_{a_i \in A_i} c_i(a_i)$
    \vspace{-0.3cm}
    \Statex
    \tikz[remember picture,overlay] {
        \node[rotate=90,anchor=south,yshift=5pt] at (0.2,-0.35) {\scriptsize \textbf{Exploitation}};
        \draw[thick] (0,-1.25) -- (0,0.4); 
    }
    \Statex \hspace{0.5em}\textbf{for} each time step $t = K+1,\ldots, T$ \textbf{do}
        \State \hspace{2.5em}play $\bar{a}_i$ 
    \State \hspace{0.5em}\textbf{end for}
    \end{algorithmic}
\end{algorithm}

We present a decentralized learning algorithm to maximize the social welfare objective under the feasibility constraint \eqref{eq: constraint}. Algorithm~\ref{alg:sbc-pe} has two phases: an exploration phase of length $K$, followed by exploitation for $T-K$ stages.

Prior work \cite{marden6426834,PRADELSKI2012882,got10.1287/moor.2020.1051} rely on internal states and aspiration levels while requiring sufficiently small design parameters for performance guarantees, without explicit bounds. In contrast, our method is  simple: it uses no state or aspiration values and only a one-bit randomized satisfaction signal, yielding an easily implementable mechanism with Pareto-optimality guarantees under explicit bounds on the design parameter $\epsilon$.

The \emph{SBC-PE} is an explore-then-commit procedure (Algorithm~\ref{alg:sbc-pe}). In the \textit{exploration phase}, agents play random actions for $K$ rounds, observe utilities, and broadcast a one-bit signal $m_i^t$ stochastically generated as in~\eqref{eq:message}. Each agent $i$ keeps counters $c_i(a_i)$, incrementing when all broadcast $\{m_j^t=1\}_{j\in N}$. After exploration, agent $i$ selects $\bar{a}_i\in\argmax_{a_i\in A_i} c_i(a_i)$. In the \textit{exploitation phase}, agent $i$ repeatedly plays $\bar{a}_i$, so the joint action $\bar{a}=(\bar{a}_1,\dots,\bar{a}_n)$ achieves the Pareto-efficient outcome. Here, each agent acts independently, using only local utility and one-bit feedback.

\section{Results}\label{sec:results}
In this section, we focus on regret-based analysis to evaluate the performance of Algorithm \ref{alg:sbc-pe}. Given the joint actions $a^t$ played at stages $t=1,\ldots,T$, the cumulative regret relative to the optimal social welfare is defined by
\begin{equation}\label{eq:regret}
   R_T = T \max_{a\in A_{\lambda}} \{W(a)\} - \sum_{t=1}^T W(a^t),
\end{equation}
where $A_{\lambda} \coloneqq \{\, a \in A : u_i(a) > \lambda_i \ \forall i \in N \,\}$ is the feasible set of joint actions.

We make the following assumptions for the regret bound:

\begin{assumption}\label{assm:1}
\begin{enumerate}[label=(\roman*), ref=\theassumption-(\roman*),noitemsep]
  \item \label{assm:1-i} For each agent, $w_i \cdot u_i(a) < 1$ for all $a \in A$.
  \item \label{assm:1-ii} There exists a unique social welfare maximizer:
  \begin{equation}
      a^* = \argmax_{a \in A_{\lambda}} \{W(a)\}.
  \end{equation}
  \item \label{assm:1-iii} For some $\delta>0$, we have 
 $\epsilon < (M+\delta)^{-1/\Delta_1}$,
 where $M \coloneqq |A_{\lambda}|$ and
 \begin{flalign}\label{eq:gap}
 &\Delta_1 \coloneqq \max_{a\in A_\lambda} \{W(a)\} - \max_{\substack{a\in A_\lambda:\\ a \neq a^*}} \{W(a)\}.
 \end{flalign}
\end{enumerate}
\end{assumption}

Define $m^t \coloneqq \prod_{j\in N} m_j^t$, taking value $1$ when each signals $m_j^t = 1$. Then, by \eqref{eq:socialwelfare} and \eqref{eq:message}, the independent randomization of messages yields
\begin{align}
P(m^t = 1 \mid a^t = a) = \left\{\begin{array}{ll} 
\epsilon^{n-W(a)} & \mbox{if } a\in A_\lambda\\
0 &\mbox{otherwise}
\end{array}\right..
\end{align}
Correspondingly, at each stage $t$, agents jointly play $a$ and signal \textit{content} with probability
\begin{equation*}
  \theta(a)\coloneqq P(a^t=a,m^ t = 1) = \left\{\begin{array}{ll}\epsilon^{n-W(a)}/|A|&\mbox{if } a\in A_\lambda\\
  0 &\mbox{otherwise}
  \end{array}\right.  
\end{equation*}
due to the uniform exploration. 

\begin{remark}\label{rem:distribution}
The content-endorsed actions are distributed by
\begin{flalign}
    P(a^t = a\mid m^t = 1) = \left\{\begin{array}{ll}
    \frac{\epsilon^{-W(a)}}{\sum_{\tilde{a}\in A_\lambda} \epsilon^{-W(\tilde{a})}} &\mbox{if } a\in A_{\lambda}\\
    0 & \mbox{otherwise}
    \end{array}\right.
\end{flalign}
\end{remark}

Since $\epsilon < 1$, Assumptions \ref{assm:1-i} and \ref{assm:1-ii} yield
\begin{flalign}
\argmax_{a \in A} \{\theta(a)\} = \argmax_{a \in A_{\lambda}} \{W(a)\} = a^*.
\end{flalign}

The following lemma shows how well $\theta(a^*)$ is separated from all others $\theta(a)$ for $a\neq a^*$.

\begin{lemma} \label{lem:bestbiggerthanall}
    Under Assumption \ref{assm:1}, we have
    \begin{flalign} \label{eq:bestbiggerthanall}
        \theta(a^*) - \xi > \sum_{\substack{a\in A_{\lambda}:\\a \neq a^*}}(\theta(a)+\xi),
    \end{flalign}
    where 
    \begin{equation}\label{eq:xi}
    \xi \coloneqq \frac{\delta}{|A|M}\cdot \epsilon^{n-\max_{\tilde{a}\neq a^*}W(a)}.
    \end{equation}
\end{lemma}

\textit{Proof:} Using Assumption \ref{assm:1}, we have $\epsilon^{-\Delta_1} > M+\delta$, and from there, we have
\begin{flalign}
    \theta(a^*)-M\xi 
    &= \frac{1}{|A|}\epsilon^{n-\max_{\tilde{a}\neq a^*}W(a)}(\epsilon^{-\Delta_1} - \delta) \\
    &> \frac{M}{|A|} \epsilon^{n-\max_{\tilde{a}\neq a^*}W(a)} > \sum_{\substack{a\in A_{\lambda}: \\ a \neq a^*}}\theta(a).
\end{flalign}
Then, by adding $(M-1)\xi$ to both sides, we obtain \eqref{eq:bestbiggerthanall}. \hfill $\square$

As an unbiased estimate of $\theta(a)$, we introduce
\begin{equation}\label{eq:thetahat}
\widehat{\theta}^t(a) \coloneqq \frac{1}{t} \sum_{k=1}^t \ind{a^k=a,m^k=1}.
\end{equation}
By the Strongly Law of Large Numbers, we have $\widehat{\theta}^t(a)\rightarrow \theta(a)$ as $t\rightarrow\infty$ almost surely for each $a\in A$. Furthermore, the marginalization yields
\begin{align}
\theta_i(\tilde{a}_i) \coloneqq \sum_{\substack{a\in A:\\ a_i=\tilde{a}_i}} \theta(a)\quad\mbox{and}\quad\widehat{\theta}_i^t(\tilde{a}_i) \coloneqq \sum_{\substack{a\in A:\\a_i=\tilde{a}_i}} \widehat{\theta}^t(a).
\end{align}

If each entry of $\theta$ can be estimated within error $\xi$, the following proposition shows that agents can locally identify the action required to achieve the optimal social welfare.

\begin{proposition} \label{prop:marginaleqjoint}
For a given stage $t$, if Assumption \ref{assm:1} holds, and $|\widehat{\theta}^t(a) - \theta(a)| < \xi$ for all $a \in A$, then we have
\begin{flalign} \label{eq:prop}
 \left(\argmax_{a_i \in A_i} \left\{\widehat{\theta}_{i}^t(a_i)\right\}\right)_{i=1}^n &= \argmax_{a \in A}\{\theta(a)\} = a^*. 
\end{flalign}
\end{proposition}

\textit{Proof:}
From the definition of $\widehat{\theta}_{i}^t$, and $\widehat{\theta}^t$, we have
\begin{flalign}
    \widehat{\theta}_{i}^t(a_i^*) = \sum_{\substack{a\in A:\\ \ a_i=a_i^*}}\widehat{\theta}^t(a)> \widehat{\theta}^t(a^*) > \theta(a^*)-\xi.
\end{flalign}
On the other hand, Lemma \ref{lem:bestbiggerthanall} yields
\begin{flalign}   
    \theta(a^*)-\xi > \sum_{a \neq a^*}(\theta(a)+\xi) &> \sum_{a \neq a^*}\widehat{\theta}^t(a)>\widehat{\theta}_{i}^t(\tilde{a}_i)
\end{flalign}
for all $\tilde{a}_i \neq a_i^*$. Hence, $\widehat{\theta}_{i}^t(a_i^*) > \widehat{\theta}_{i}^t(\tilde{a}_i)$ for all $\tilde{a}_i \neq a_i^*$, and we obtain \eqref{eq:prop}. \hfill $\square$

\begin{remark}
The proof of Proposition \ref{prop:marginaleqjoint} clarifies the need for Assumption \ref{assm:1-ii}. With multiple maximizers, Lemma \ref{lem:bestbiggerthanall} fails, leaving agents unable to choose the socially better action, and the resulting joint action may be arbitrarily worse.
\end{remark}

\begin{remark}
If the joint action is observable to all agents, the dynamics still guarantee convergence to one of the social maximizers without Assumptions \ref{assm:1-ii} and \ref{assm:1-iii}.
\end{remark}

The following theorem provides $O(\log(T))$ bound on the expected regret \eqref{eq:regret}, where the expectation is taken with respect to the randomness induced by the uniform exploration and stochastic feedback.

\begin{theorem}{Main Result} \label{thm:main}
For any $n$-agent game, if Assumption \ref{assm:1} holds and the exploration length is $K=\tfrac{\log(4MT\xi^2)}{2\xi^2}$, then the expected total regret is bounded by
\begin{flalign}\label{eq:regretbound}
    \mathrm{E}[R_T] < \frac{\Delta}{2\xi^2}\left(1+\log(4MT\xi^2)\left(1-\frac{1}{2T\xi^2}\right)\right),
\end{flalign}
where $\Delta \coloneqq \max_a (W(a^*)-W(a))$, $M=|A_\lambda|$, and $\xi$ is as described in \eqref{eq:xi}.
\end{theorem}

\textit{Proof:}
Based on the randomness on $\mathbb{I}_{\{a^t=a,m^t=1\}}\in \{0,1\}$ and \eqref{eq:thetahat}, the Hoeffding bound yields that
\begin{flalign}\label{eq:hoeffding}
P\left(\left|\widehat{\theta}^K(a)-\theta(a) \right|\geq \xi \right) \leq 2e^{-2K\xi^2}
\end{flalign}
for each $a$. Then, using the union bound, we obtain
\begin{flalign*} 
P\left(\left|\widehat{\theta}^K(a)-\theta(a)\right|\geq \xi\mbox{ for some } a\in A\right) \leq 2Me^{-2K\xi^2}.
\end{flalign*}
Correspondingly, we have
\begin{flalign}\label{eq:lambda}
&P\left(\left|\widehat{\theta}^K(a)-\theta(a)\right| \leq \xi, \ \forall a \in A\right) \geq 1- \beta,
\end{flalign}
where 
$\beta \coloneqq 2Me^{-2K\xi^2}$.   
By Proposition~\ref{prop:marginaleqjoint} and \eqref{eq:lambda}, agents play the best joint action $a^*$ in the exploitation phase with probability at least $1-\beta$ and a worse joint action $a\neq a^*$ with probability at most $\beta$. Thus, the expected regret is:
\begin{flalign*}
    \mathrm{E}[R_T] \leq K\Delta + \beta (T-K)\Delta < K\Delta + \beta T \Delta.
\end{flalign*}
The optimal $K^*$ minimizing $K\Delta + \beta T \Delta$ is given by
\begin{flalign*}
    K^* = \tfrac{1}{2\xi^2} \log(4MT\xi^2).
\end{flalign*}
Choosing $K = K^*$, we obtain $\beta = \frac{1}{2T\xi^2}$ and \eqref{eq:regretbound}, which completes the proof. \hfill $\square$

\section{Simulation}
\label{sec:simulation}
\begin{figure}[t]
    \centering
    \includegraphics[width=.95\linewidth]{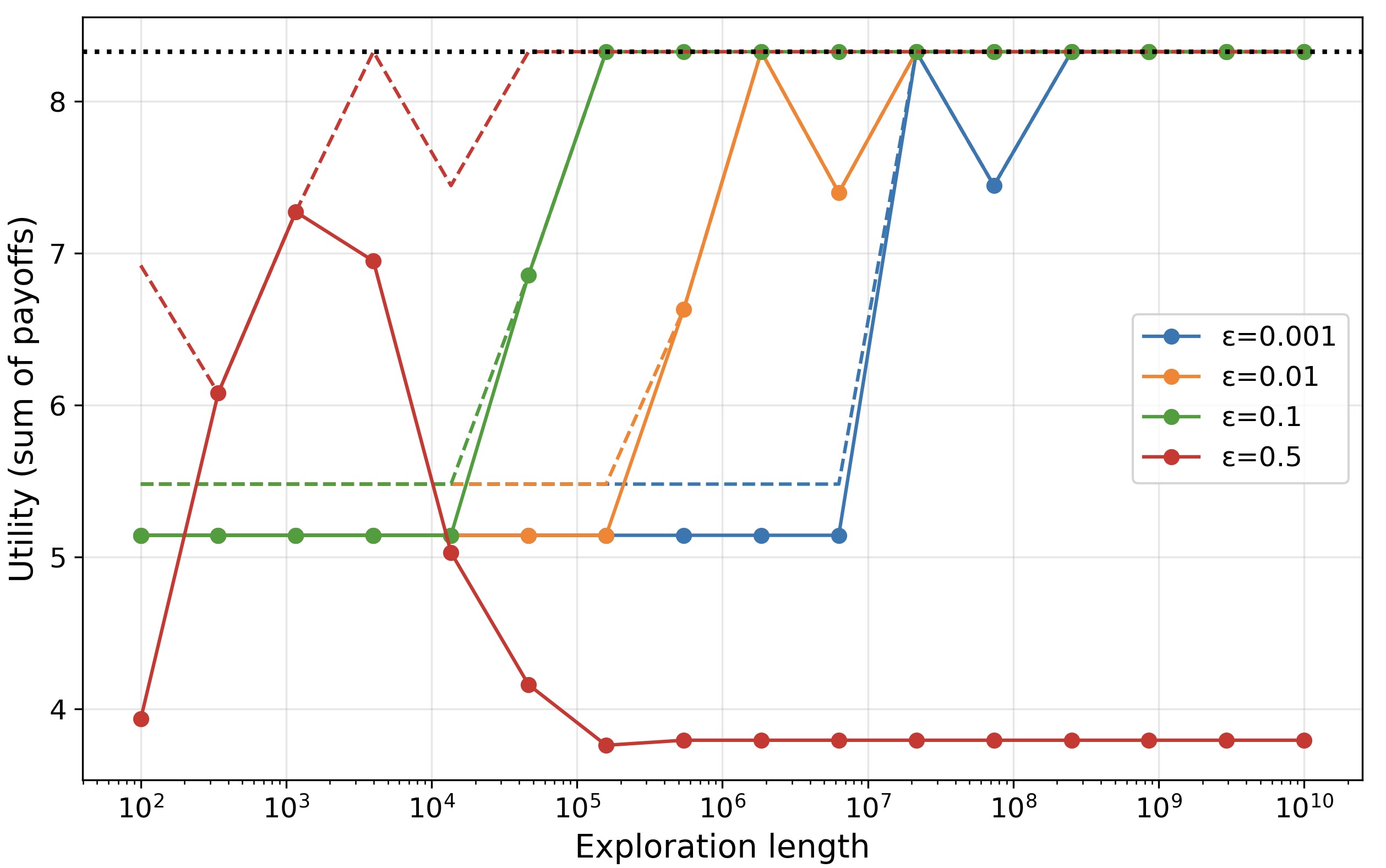}
    \caption{Performance of Algorithm~\ref{alg:sbc-pe} for $n=10$ agents. 
The x-axis is the exploration length $K$ (log scale), and the y-axis the total utility. 
Solid lines depict the committed joint action $\bar{a}$, dashed lines the most-frequently-played content-endorsed joint action, and the black dotted line the optimal utility.}
    \label{fig:simulation_results_n10_m2_018}
\end{figure}

We examine the performance of the 
proposed SBC-PE algorithm for different exploration lengths $K$, $\epsilon$ parameters and the welfare gap $\Delta_1$, as described in \eqref{eq:gap}. To this end, we focus on $(n=10)$-agent games in which each agent $i\in N$ has two actions, $w_i=1$, and $\lambda_i=0.2$. We set utilities $u_i(a)$ randomly over $[0,1]$, consistent with \eqref{eq: constraint}.



In Fig.~\ref{fig:simulation_results_n10_m2_018}, we examine the impact of $K$ and $\epsilon$ while comparing the performances of committed joint action and the most-frequently-played content-endorsed action, i.e., $\argmax_a \widehat{\theta}(a)$. This illustrates that larger $\epsilon$ values can hinder efficiency while smaller $\epsilon$ values can require longer exploration. Note that Assumption \ref{assm:1-iii} provides an explicit bound on $\epsilon$ ensuring Pareto-efficient outcomes for sufficiently large $K$, as shown in Theorem \ref{thm:main}. In this specific game, the decentralized algorithm of~\cite{marden6426834} fails to stabilize at the Pareto-efficient outcome. Even with the best choice of $\epsilon\in[10^{-5},0.5]$ over $10^{10}$ steps, the algorithm spends only a negligible fraction of time at the social optimum, exhibiting an average welfare gap of approximately $11\%$ relative to $\max_{a\in A_{\lambda}} W(a)$. 



\begin{figure}[t]
    \centering
    \includegraphics[width=0.95\linewidth]{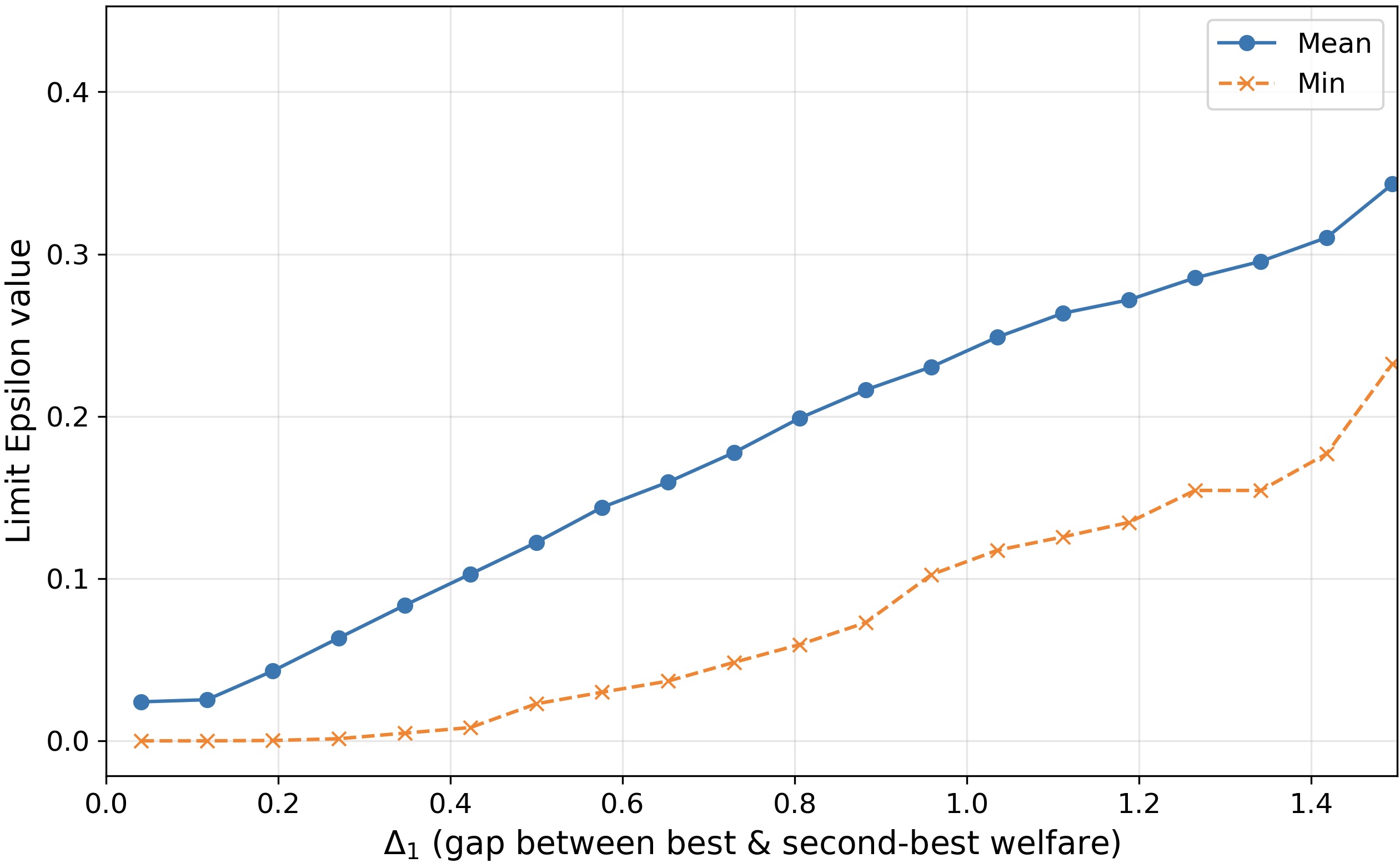}
    \caption{Limit values of $\epsilon$ as a function of the welfare gap $\Delta_1$ for $n=10$ agents and two actions. The blue curve indicates the mean of the limiting $\epsilon$ values, while the orange curve shows the minimum acceptable $\epsilon$.}    
    \label{fig:delta_vs_eps_n10_m2}
\end{figure}

In Fig.~\ref{fig:delta_vs_eps_n10_m2},
we further examine the average and the minimum $\epsilon$ parameters ensuring Pareto-efficient outcomes across $20{,}000$ game samples for different welfare gaps $\Delta_1$. For large $\Delta_1$, the frequency of the content-endorsed socially maximizing action increases (See Remark \ref{rem:distribution}). This allows larger $\epsilon$ values, showing the robustness of SBC-PE when the optimal solution is well separated from suboptimal ones.

\section{Discussion \& Conclusion}\label{sec:conclusion}

Simulations confirm that SBC-PE achieves exact socially optimal outcomes with minimal communication. The parameter $\epsilon$ is critical: too small leads to slow adaptation, while too large reduces efficiency. The exploration horizon $K$ must also scale with system size to ensure reliable identification of Pareto-efficient profiles.

\begin{remark}
Consider that agents can observe the full joint action and use the message rule 
\begin{equation}
\label{eq:message2}
    m_i^t =
    \begin{cases}
        \ind{a_i^t \in \mathrm{BR}_i(a_{-i}^t)} & \textrm{with probability } \epsilon^{\,1 - w_i \cdot u_i^t}, \\
        0 & \textrm{otherwise},
    \end{cases}
\end{equation}
where 
$\mathrm{BR}_i(a_{-i}^t) \coloneqq \argmax_{a_i\in A^i} \{u^i(a_i,a_{-i}^t)\}$ is the best response set. With \eqref{eq:message2}, Algorithm~\ref{alg:sbc-pe} can learn Pareto-Optimal Pure Equilibrium if a pure equilibrium exists.
\end{remark}

Communication-based methods guaranteeing Pareto optimality often face scalability challenges from heavy message exchange. SBC-PE remains efficient since each agent broadcasts only one bit per stage, independent of the game size. Thus, the overall communication cost grows linearly with the number of agents while message size is fixed. Hence, the algorithm is communication efficient and scalable. SBC-PE demonstrates that strong welfare optimization is possible even under very limited communication. This makes it a practical candidate for large-scale applications.

\bibliographystyle{IEEEbib}
\bibliography{refs}

\end{document}